\documentclass[
reprint,
superscriptaddress,
showpacs,
nofootinbib,
amsmath,amssymb,
aps,
prc,
]{revtex4-1}

\usepackage{graphicx}
\usepackage{dcolumn}
\usepackage{bm}
\usepackage{multirow}

\usepackage{hyperref}

\begin{document}


\title{Cross sections for neutron-deuteron elastic scattering in the energy range 135-250~MeV}



\author{E.~Ertan}
\email[]{erol.ertan@boun.edu.tr}
\affiliation{Department of Physics, Bogazici University, 34342 Istanbul, Turkey}

\author{T.~Akdogan}
\affiliation{Department of Physics, Bogazici University, 34342 Istanbul, Turkey}
\affiliation{Department of Physics and Laboratory for Nuclear Science,
Massachusetts Institute of Technology, Cambridge, MA 02139}

\author{M.~B.~Chtangeev}
\altaffiliation[Present Address: ]{Aurora Flight Sciences, 
Cambridge, MA 02142}
\affiliation{Department of Physics and Laboratory for Nuclear Science,
Massachusetts Institute of Technology, Cambridge, MA 02139}

\author{W.~A.~Franklin}
\altaffiliation[Present Address: ]{Passport Systems Inc., Billerica, MA 01862}
\affiliation{Department of Physics and Laboratory for Nuclear Science,
Massachusetts Institute of Technology, Cambridge, MA 02139}

\author{P.~A.~M.~Gram}
\altaffiliation[Present Address: ]{9 Wagon Meadow Road, Santa Fe, NM 87505}
\affiliation{Los Alamos Neutron Science Center, Los Alamos National Laboratory,
Los Alamos, NM 87545}

\author{M.~A.~Kovash}
\affiliation{Department of Physics and Astronomy, University of Kentucky, Lexington, KY 40506}

\author{J.~L.~Matthews}
\affiliation{Department of Physics and Laboratory for Nuclear Science,
Massachusetts Institute of Technology, Cambridge, MA 02139}

\author{M.~Yuly}
\affiliation{Department of Physics, Houghton College, Houghton, NY 14744}


\date{\today}

\begin{abstract}
We report new measurements of the neutron-deuteron elastic scattering cross
section at energies from 135 to 250 MeV and center-of-mass angles from
$80^\circ$ to $130^\circ$. Cross sections for neutron-proton elastic
scattering were also measured with the same experimental setup for
normalization purposes. Our $nd$ cross section results are compared with
predictions based on Faddeev calculations including three-nucleon forces,
and with cross sections measured with charged particle and neutron beams
at comparable energies.
\end{abstract}



\maketitle


\section{Introduction \label{intro}}

Three-nucleon systems comprise an area of long-standing interest in
nuclear physics. The study of these systems has been enhanced by the
growing database of precise measurements on two-nucleon systems~\cite{Arnd_2000},
and the ability of modern potential models to provide accurate predictions
of nucleon-nucleon scattering observables~\cite{Stok_1993, Wiri_1995,
bib:machleidt2001,bib:deltuva2003}. Furthermore, modern computational
techniques have made it possible to calculate scattering cross sections
and spin observables in three-nucleon systems for any kinematical configuration
using the Faddeev formalism~\cite{Gloc_1996}, allowing the identification of
experiments with strong sensitivity to the effects of three-nucleon forces (3NF).
A comprehensive review of the theoretical and experimental status with regard to
3NF in few-nucleon systems has recently been published by Kalantar-Nayestanaki,
Epelbaum, Messchendorp, and Nogga~\cite{Kala_2012}.

In 1998, Wita\l{a} and co-workers~\cite{Wita_1998} employed Faddeev techniques
to compute differential cross sections for $nd$ elastic scattering at 65, 140,
and 200 MeV, using solely two-nucleon forces as well as with the inclusion of
the Tucson-Melbourne (TM) three-nucleon force~\cite{Coon_1981}. The cross
section is seen to fall steeply with angle, due to the direct term, and then
to rise again at backward angles due to the exchange term in the two-nucleon
interaction. The 3NF term alone is nearly isotropic and thus is of magnitude
comparable to or greater than that of the two-nucleon term for angles at which
the cross section exhibits a minimum. Similar results were achieved by Deltuva,
Machleidt and Sauer~\cite{bib:deltuva2003} in which a coupling of
nucleon-$\Delta$-isobar states to two-nucleon states is considered. Their study
demonstrated that this coupling gives rise to an effective 3NF force and a pion
exchange mechanism. This suggests that measurements of the intermediate-to-large
angle $Nd$ cross section could reveal effects of 3NF. Moreover, the 3NF predictions
showed significant variation with incident energy, indicating that energy-dependent
measurements would be valuable.

In recent years, several measurements have been carried out in this kinematic
region, supplementing early measurements of $nd$ cross sections near 150
MeV~\cite{bib:Palmieri1972} and $pd$ cross sections near both
140~MeV~\cite{bib:Postma1961, Igo_1972} and 200~MeV~\cite{PhysRevD.5.2139, Igo_1972}.

Measurements of differential cross sections for $dp$ elastic scattering at
$E_d$ = 270 MeV performed at the RIKEN Accelerator Research Facility were
reported in 2000 by Sakai {\it et al.}~\cite{bib:Sakai_2000} and in 2002 by
Sekiguchi {\it et al.}~\cite{Seki_2002}. Measurements of $pd$ elastic
scattering cross sections have been performed at six energies between 108 and
190~MeV at the KVI facility~\cite{Ermi_2003, Ermi_2005}. The cross sections
at $E_p$ = 135 MeV were found to be 10-40\% larger than those measured at RIKEN.
Subsequent measurements by Sekiguchi {\it el al.}~\cite{Seki_2005} at 135~MeV/A
using both proton and deuteron beams support the original RIKEN measurements and
contradict those from KVI. In 2008, Ramazani-Moghaddam-Arani {\it et al.}~\cite
{bib:Ramazani2008} reported a new measurement of the $pd$ cross section at KVI
which yielded a result intermediate between the earlier KVI and the RIKEN cross
sections.

There have been fewer studies of the $nd$ cross section than of the $pd$ cross
section. Mermod {\it et al.}~\cite{Merm_2004, Merm_2005} measured the differential
cross section for $nd$ scattering at 95~MeV, and Maeda {\it et al.}~\cite{Maed_2007}
performed a measurement for center-of-mass angles from $10^{\circ}$ to $180^{\circ}$
at $E_n$ = 248 MeV. These data were found to agree well with the previous $pd$ cross
section measurements at $E_p$ = 252 MeV reported in 2002 by Hatanaka
{\it et al.}~\cite{Hata_2002}.

Many of the previous experiments have been performed with polarized beams, and
analyzing powers and other spin observables have been reported~\cite{Ermi_2001,
Seki_2002, Hata_2002, Ermi_2005, PhysRevC.74.064003, Amir_2007,Maed_2007, Step_2007,
bib:Ramazani2008, Seki_2009, Seki_2011}. Discussion of these, along with measurements
of the inelastic $Nd$ (breakup) cross section~\cite{Kist_2003, Kist_2005, Step_2010,
Ciep_2012}, are beyond the scope of this paper.

As stated earlier, most of the previous experiments have been carried out with
charged particle beams, necessitating the consideration of Coulomb effects on
the cross section. Also, with the exception of the KVI work~\cite{Ermi_2003,
Ermi_2005} and the early work of Igo {\it et al.}~\cite{Igo_1972}, previous
measurements have been performed at a single energy. Moreover, there are still
some lingering uncertainties in the magnitude and shape of the differential cross
section at 135 MeV~\cite{bib:Ramazani2008}.

These factors have motivated the present measurement: a study of the
neutron-deuteron elastic scattering cross section, at large angles where the
sensitivity to 3NF is greatest, over a broad range of incident neutron energies.

Since the absolute normalization of neutron scattering experiments can be a
difficult problem, the $np$ scattering cross section was also measured using
the same experimental setup. The $np$ data were then used to normalize the
measured $nd$ cross sections. Other recent $nd$ measurements have used this
technique to achieve precision sufficient to distinguish among calculations
which display the explicit effects of 3NF~\cite{Merm_2005}.

\section{Experiment \label{exp}}

The experiment was carried out at the Los Alamos Neutron Science Center (LANSCE)
at the Los Alamos National Laboratory, Los Alamos, New Mexico. Neutrons were
produced as spallation products from an 800 MeV H$^-$ beam incident on a bare
tungsten target. H$^-$ pulses from the linear accelerator had a width of
approximately 0.2~ns, with an inter-pulse separation of 1.8~$\mu$s. This pulse
structure allowed the energy of outgoing (and incident) neutrons to be measured
by time-of-flight techniques. The spallation method produced a ``white" source
of neutrons ranging from very low energies up to nearly 800 MeV, with an energy
distribution that depends on the production angle.

This experiment was performed on a forward-angle, $15^\circ$ with respect to the
primary H$^-$ beam, flight path, to maximize the flux of neutrons in the energy
regime above 100 MeV. The neutron beam was defined by two sets (horizontal and
vertical) of Cu shutters with typical aperture of 3.8 $\times$ 3.8 cm, followed by a
sweep magnet to filter out charged particles. The beam entered an evacuated pipe
containing a 2.7~m long steel collimator with an 1.3-cm diameter circular aperture
surrounded by magnetite shielding, and exited to pass through a ${}^{238}$U foil
fission ionization chamber~\cite{Wend_1993} that monitored the beam flux as a
function of neutron energy. Approximately one meter downstream from the fission
chamber, the beam impinged on a cryogenic target cell containing either liquid
deuterium (LD$_2$) or liquid hydrogen (LH$_2$). The target was located approximately
18~m from the spallation source. The intensity profile and position of the beam
were measured at the target position by exposing a storage-phosphor image plate.

The target geometry and orientation were designed to minimize the energy loss and
reactions of outgoing charged particles. The target cell consisted of a horizontal
cylindrical disk of 1.3~cm thickness and 12.7~cm diameter, with 51~$\mu$m Aramica
entrance and exit windows. It was placed at an angle of $50^\circ$ with respect to
the incident neutron beam to provide an optimum path for the outgoing charged particles
and to eliminate mechanical constraints for neutrons exiting the cell at large angles
on the opposite side of the beam. The target, located in an evacuated cylindrical
scattering chamber with a 127~$\mu$m Kapton window, was cooled using a cryogenic
refrigeration system which employed gaseous $^4$He as its working fluid. The system
had a nominal cooling capacity of 10~W at 20~K; a resistive heater was used to
maintain a nominal absolute target pressure of 97~kPa.

Scattered neutrons and recoiling charged particles were observed in coincidence.
Protons and deuterons were detected by five telescopes, each consisting of a thin
$\Delta E$ plastic scintillator backed by a pure CsI calorimeter. These detectors
were positioned with their front faces 100~cm from the center of the target, at
mean laboratory angles of $\theta_{\mathrm{lab}} = 24^\circ$, 30$^\circ$, 36$^\circ$,
42$^\circ$ and 48$^\circ$. The $\Delta E$ scintillators were 0.25 cm in thickness,
and provided accurate information on the particle arrival time with an efficiency
close to 100\%. Their active areas of $10\times 10$~cm$^2$ defined the solid angle
for charged particle detection. The CsI calorimeters were 30~cm in depth, and
provided a measure of both the particle energy and its arrival time.

\begin{figure*}[hbt!]
  \centering
  \includegraphics[width=0.9\textwidth]{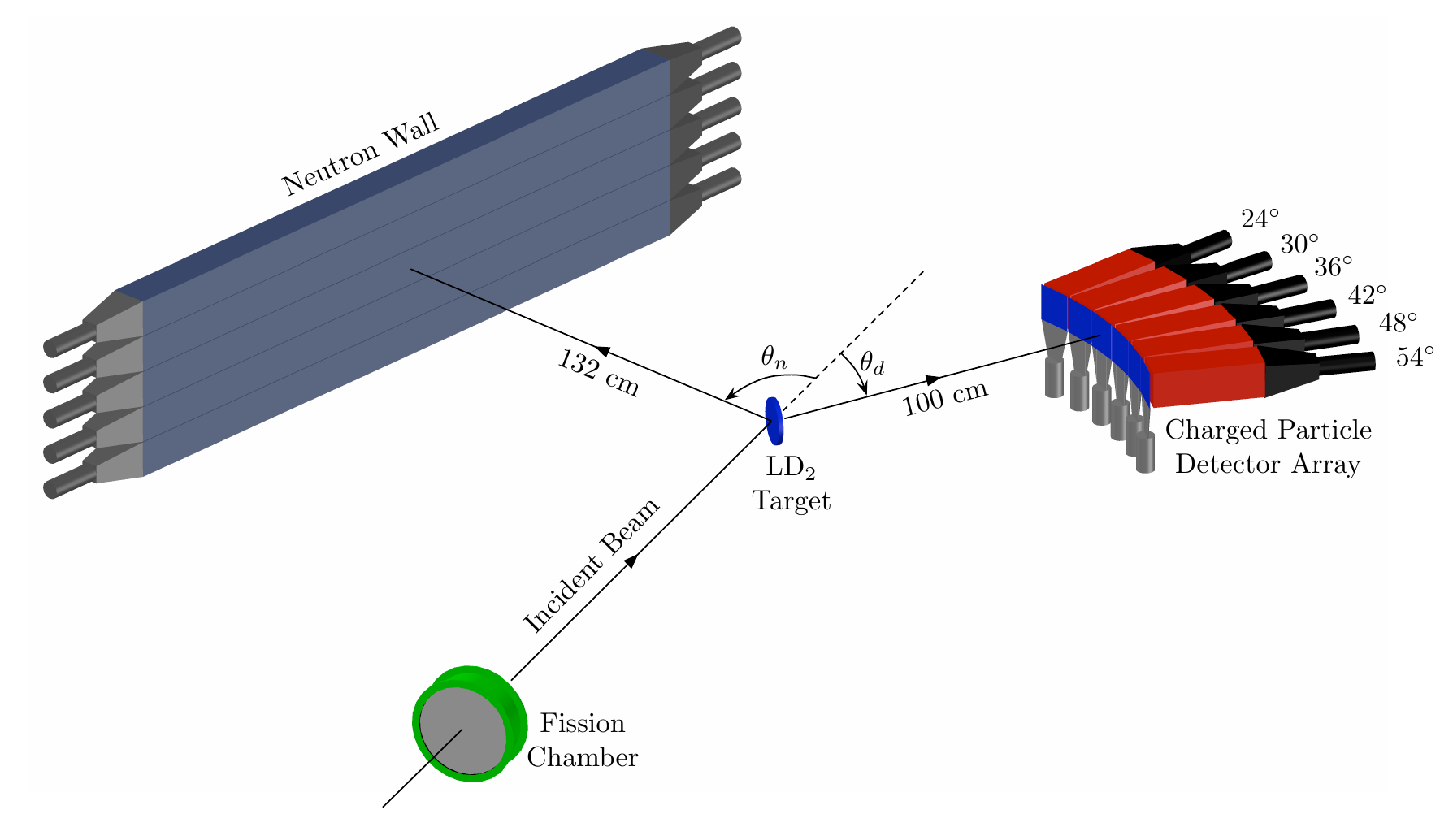}
  \caption{Schematic diagram of the experimental setup.}\label{fig:setup}
\end{figure*}

The scattered neutrons were detected with five plastic scintillator bars, each
10~cm high $\times$ 10~cm thick $\times$ 200~cm wide. The bars were stacked
vertically to form a ``wall" 200~cm wide $\times$ 50~cm high. Photomultiplier
tubes were attached to the ends of each bar to allow both the neutron hit position
and its time-of-flight to be determined. The center of the neutron wall was positioned
at a distance of 1.32~m from the target, and spanned a laboratory angle range from
34$^\circ$ to 108$^\circ$. The face of the neutron wall was covered by four thin
plastic veto scintillators to eliminate events produced by charged particles. The
energy threshold of each bar was determined with low energy gamma-ray sources,
and the photomultiplier gains were continuously monitored using cosmic-ray triggers.
The experimental setup is illustrated in Figure~\ref{fig:setup}. For $nd$ elastic
scattering, the average neutron angles in the laboratory and center of mass systems
corresponding to each recoil deuteron detector are given in Table~\ref{tab:angles}.

\begin{table}[h]
  \caption{The average angles in lab and center-of-mass frames. }\label{tab:angles}
  \centering
  \begin{tabular}{| c |  c | c |}
  \hline
  $\theta_d$ (Lab) &  $\theta_n$ (Lab)  & $\theta_n$ (CM) \\
  \hline\hline
  24  &  100  &  131  \\
  \hline
  30  &  86  &  119  \\
  \hline
  36  &  75  &  107  \\
  \hline
  42  &  65  &  95  \\
  \hline
  48  &  56  &  83  \\
  \hline
  \end{tabular}
\end{table}

Data acquisition was performed using standard NIM and CAMAC modules. A pre-scaled
fraction of the ``singles" counts in each detector arm was read out by the electronics
in addition to the coincidence events. Empty-target runs were interspersed throughout
the experiment to provide a measure of background. The target was filled with LD$_2$
for the $nd$ elastic scattering cross section measurements and with LH$_2$ for
normalization purposes. The experiment has been described in detail in~\cite{bib:Max2005}.

\section{Data Analysis \label{anal}}

\begin{figure}[hbt!]
  \centering
  \includegraphics[scale=0.24]{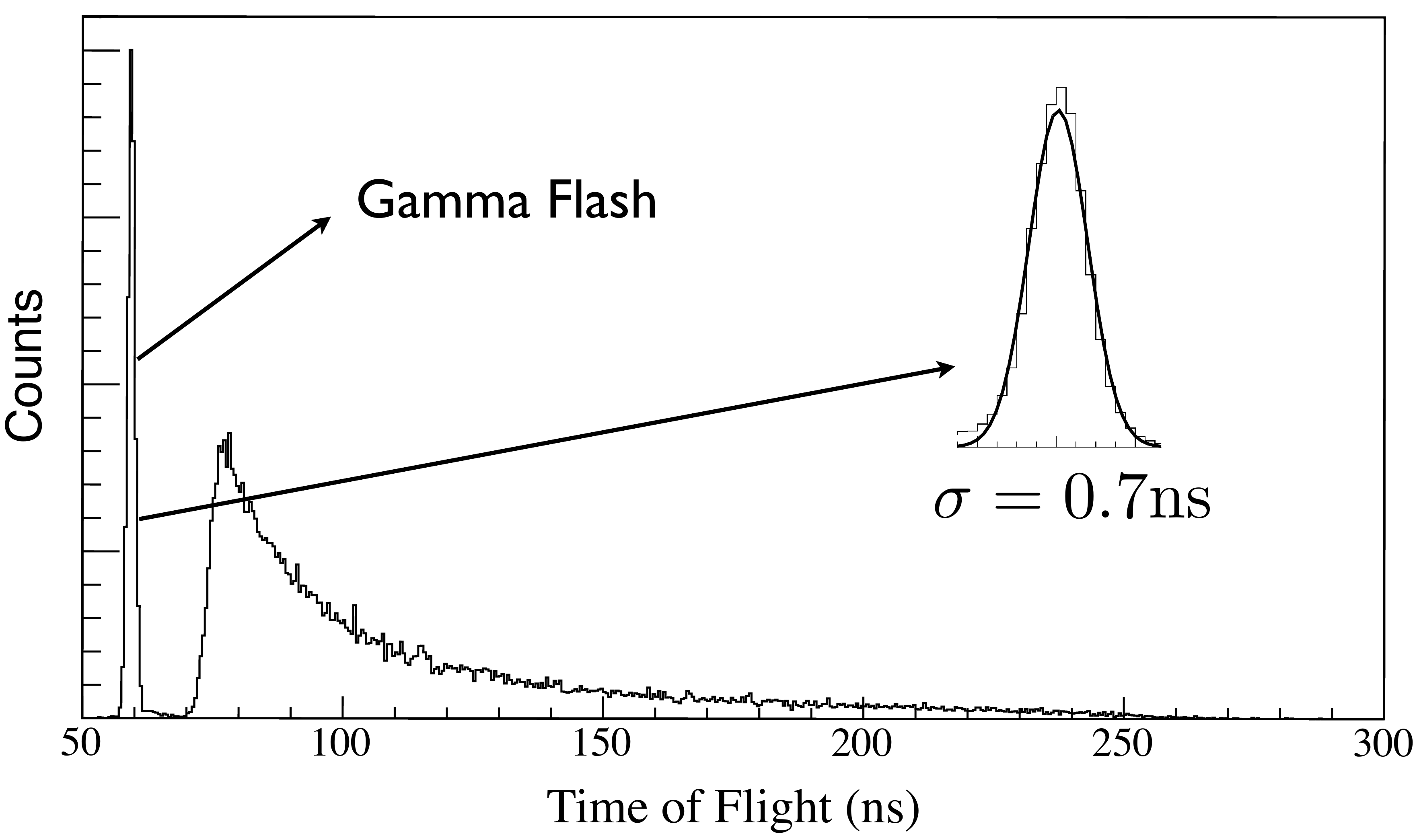}
  \caption{Typical time-of-flight spectrum for a $\Delta E$ detector, measured with
  respect to the arrival time of the proton burst at the spallation target ($t_0$).
  The``gamma flash" is observed at 60~ns. Charged particles are seen between 70 and
  250~ns.}\label{fig:gamma}
\end{figure}

Identification of $np$ and $nd$ elastic scattering events was achieved using a
succession of cuts based on the charged particle and neutron detector pulse height
and time information. The sequence began with a cut on the neutron beam energy.
When the proton beam impinged on the tungsten spallation target, a time reference
signal, called $t_0$, was generated. Since gamma rays as well as neutrons and
charged particles were produced in the spallation target, the ``gamma flash" was
easily identifiable as the leading peak in the time-of-flight spectra such as that
shown in Figure~\ref{fig:gamma}. The position of this peak could be used to establish
the time offset calibration for each detector with an uncertainty of $\pm$0.67 ns
(stat) $\pm$0.20 ns (sys). The neutron beam energy for each event trigger was then
deduced using the measured time between $t_0$ and the trigger observed in the
$\Delta E$ scintillator. This calculation assumes elastic scattering kinematics,
and includes a small time correction determined by Monte Carlo simulation for the
charged particle energy loss. The uncertainty in incident energy using this method
was estimated to be 2~MeV. Scattering cross sections were ultimately extracted
in bins of 10~MeV width.

\subsection{Neutron-proton Scattering Analysis}\label{sec:np}

Elastic neutron-proton scattering, using a LH${}_2$ target, was observed primarily
in order to determine the neutron beam flux and the target thickness. Although both
of these quantities were measured, several factors contributed to the uncertainty
in these measurements. The fission chamber was well calibrated for neutron energies
below 100 MeV, but uncertainties in the ${}^{238}$U fission cross section limited
the accuracy of the calibration at higher energies. The physical thickness of the
cryogenic target could be measured precisely at atmospheric pressure and room
temperature, but not when the target was under vacuum and filled with liquid
hydrogen or deuterium. Visual inspection of the target through the window of
the scattering chamber under operating conditions revealed two effects, a
bulging-outward of the cell windows and a steady stream of rising bubbles, both
of which would change the effective target thickness. 

\begin{figure}[hbt!]
  \centering
  \includegraphics[scale=0.8]{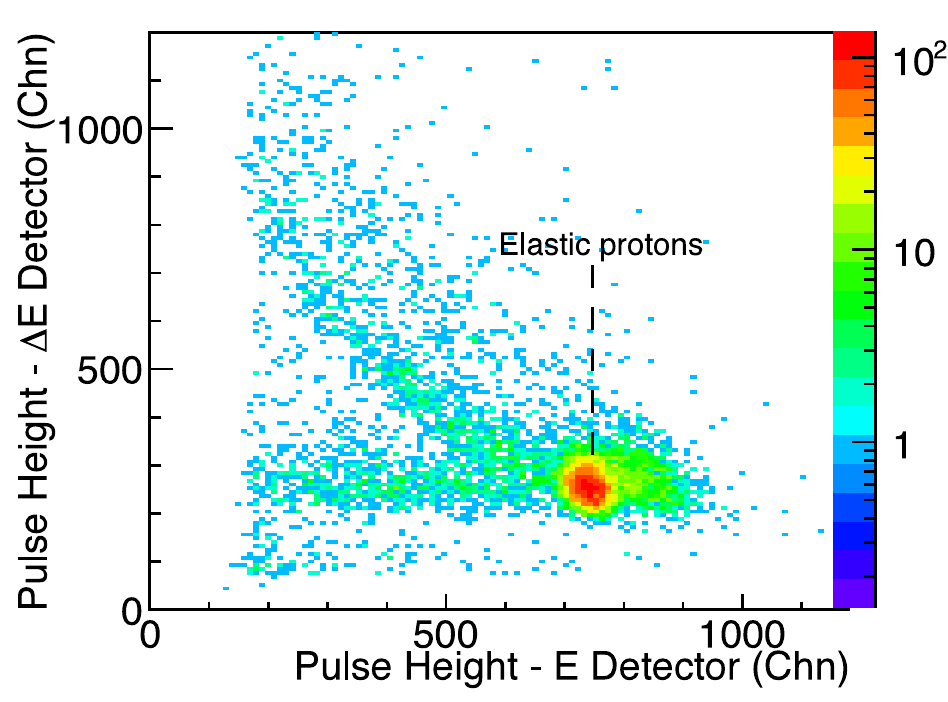}
  \caption{$\Delta E$ plotted versus $E$ for $np$ scattering at incident neutron
  energy 170~MeV and proton recoil angle 36$^{\circ}$. The peak in the correlated
  (marked) band due to protons is clearly visible. Protons of other energies can
  arise from elastic (higher energy) and inelastic (lower energy) scattering from, e.g.,
  carbon and oxygen nuclei in the windows of the target and scattering chamber.
  The horizontal band represents elastically scattered protons that have suffered
  strong-interaction energy losses in the CsI. These are included as valid events.}
  \label{fig:EdE_np}
\end{figure}

After the cut on the neutron beam energy, graphical cuts were applied on the
$\Delta E-E$ histograms for the five charged particle telescopes. An example is
shown in Figure~\ref{fig:EdE_np}. Since the event rate was dominated by $np$ elastic
scattering, the recoil protons clearly stand out over backgrounds from protons
elastically and inelastically scattered by other materials (curved band) and from
protons experiencing strong interactions in the detectors as well as particles other
than protons (horizontal band). For the data taken with hydrogen, these cuts were
sufficiently selective to provide a good measure of the efficiency for charged
particle detection, even when using the proton singles trigger in which information
from the neutron detectors was not considered. This was particularly true after the
backgrounds, determined from data taken with the target evacuated, were subtracted.

\begin{figure}[hbt!]
  \centering
  \includegraphics[scale=0.32]{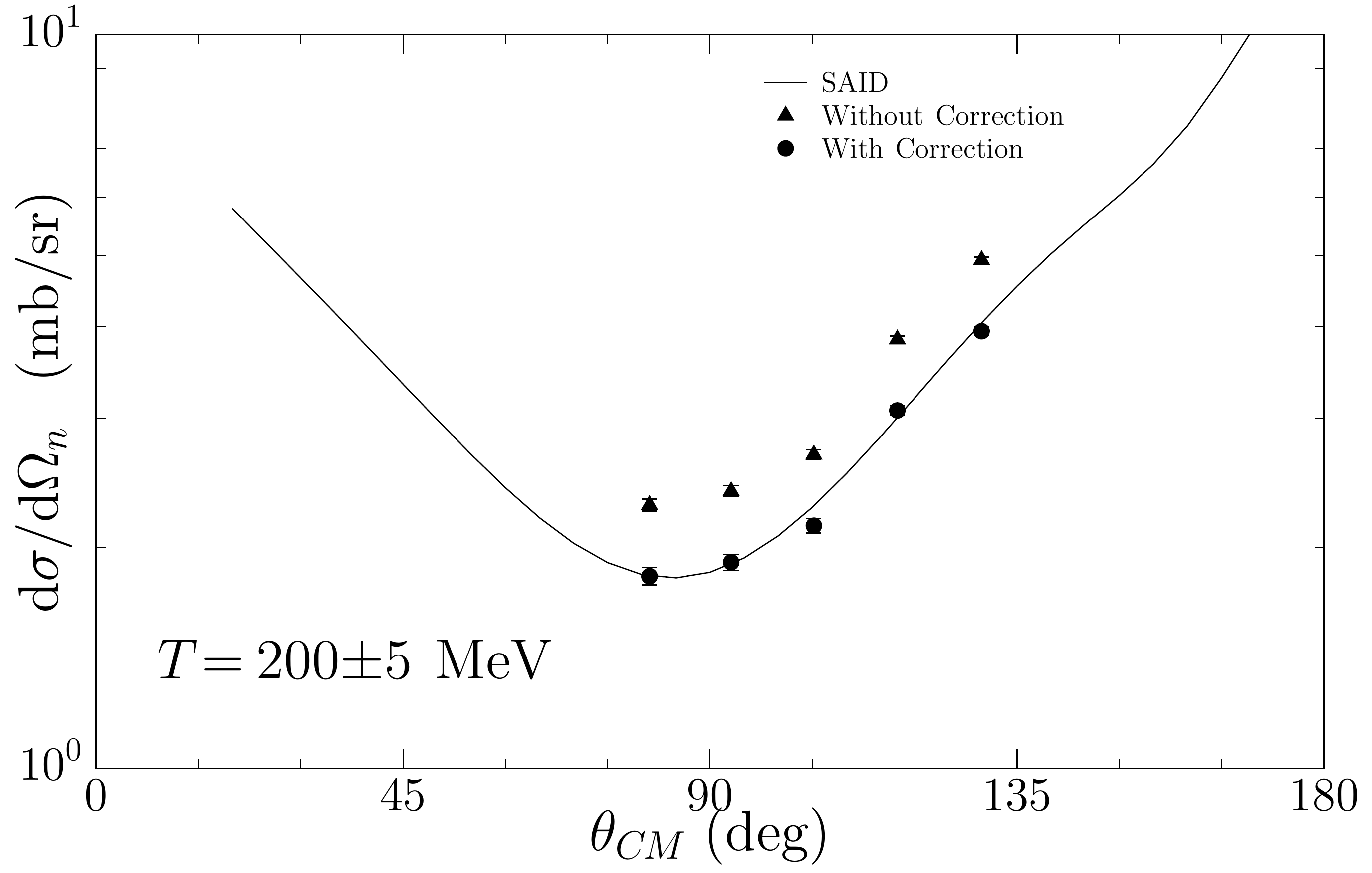}
  \caption{The $np$ cross section results with and without the correction
  factor of 1.25 $\pm$ 0.06.}
  \label{fig:np_norm}
\end{figure}

The measured $np$ scattering cross sections, with a ``floating" absolute
normalization, were compared to predictions from the SAID multi-energy partial
wave analysis~\cite{Arnd_2000}, known to be accurate at the 1\% level in
this energy regime based on comprehensive fits to the SAID $NN$ database.
Use of alternative partial wave analyses produced little difference in the results.
Target thickness being the dominant factor in this renormalization, no strong
energy dependence of the renormalization factor is expected; hence 
an overall renormalization factor of 1.25 was determined and was subsequently
applied to the $nd$ scattering data. Figure~\ref{fig:np_norm} shows the $np$
cross section obtained from this work before and after renormalization.

Assuming the fission chamber calibration to be substantially correct, the
correction factor represents a $\sim$25\% increase in the effective target
thickness.\footnote{With regard to the validity of applying the same renormalization
factor to the $nd$ data: the fission chamber calibration and the window-bulging will
certainly be the same for both hydrogen and deuterium. Following the experiment, the
target cell was observed (at room temperature and atmospheric pressure) to have a
permanent deformation; its physical thickness had increased by a factor of 1.45.
The reduction of the effective target thickness due to bubbling apparently leads to
the factor 1.25. In principle the reduction could be different for LH${}_2$ and
LD${}_2$, since the boiling point temperature for LD${}_2$ is lower. However, the
systematic error associated with the assumption of a constant renormalization
factor is estimated to be 10\% at worst, and has been included in the analysis.}

\subsection{Neutron Detector Efficiency}

\begin{figure}[hbt!]
  \centering
  \includegraphics[scale=0.55]{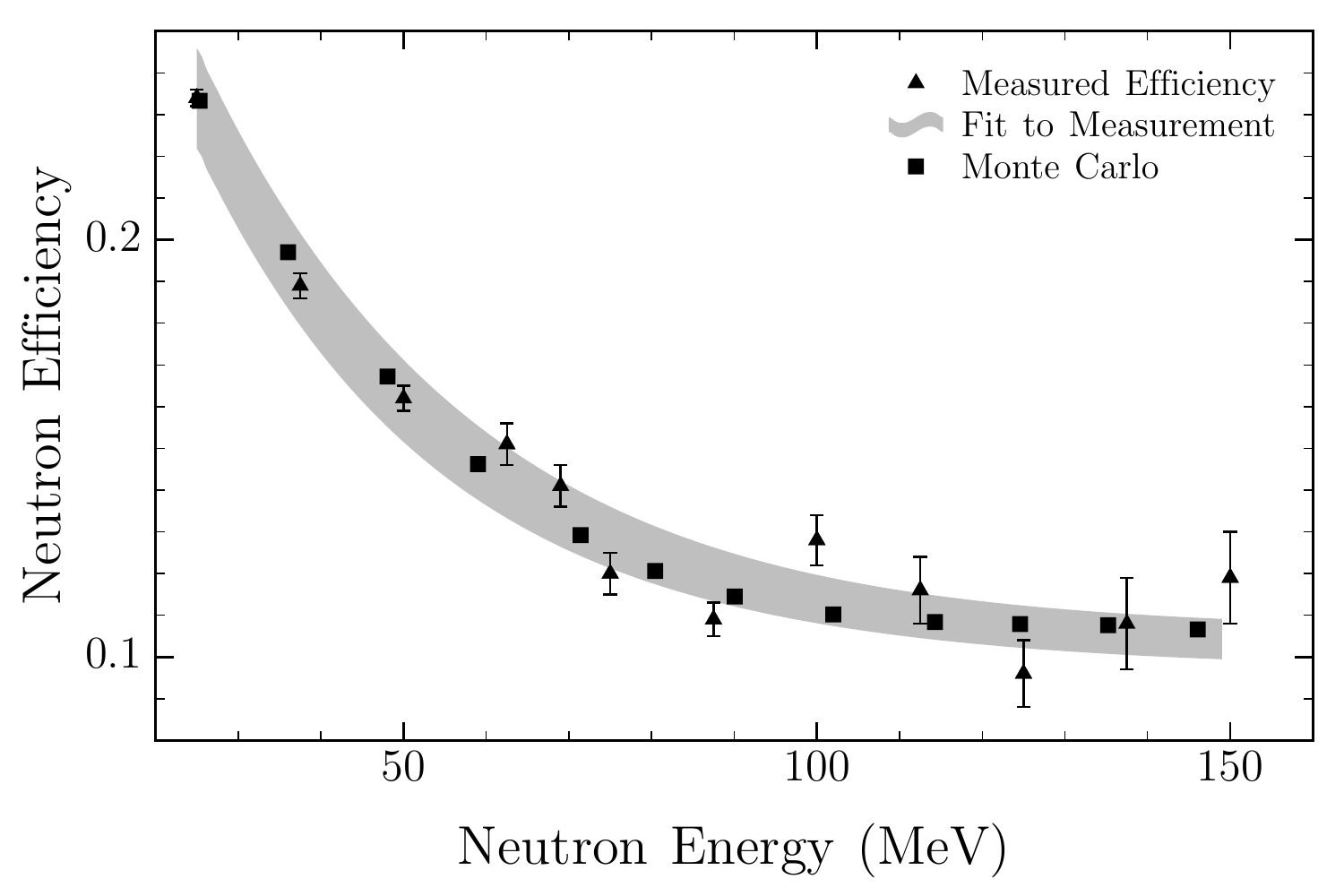}
  \caption{Neutron detection efficiency in the scintillator bars as a function
  of neutron kinetic energy. The triangles represent the data obtained in the present
  measurement, the squares a Monte Carlo simulation \cite{bib:Cecil1979}. The band
  is a fit to the data as described in the text.}
  \label{fig:neff_band}
\end{figure}

In addition to providing the normalization, the $np$ scattering data were used
to determine the efficiency of the neutron detectors. This is necessary since
the $nd$ data analysis requires the neutron-charged particle coincidence trigger.
Scattered neutron energies were determined by the measured beam energy and the
proton scattering angle. The ratio of the number of coincident neutrons observed
in the relevant regions of the wall of neutron bars to the total number of measured
protons was used to determine the neutron detector efficiency. The results are
presented in Figure~\ref{fig:neff_band} as a function of neutron energy. Also shown
in the figure are the results of a Monte Carlo simulation~\cite{bib:Cecil1979}. 
The band shows the range of efficiencies which were found by fitting the data
with a decaying exponential. Due to the effect of light attenuation in the
scintillator, the efficiency decreases for neutrons incident near the ends
of the bar. To account for this position dependence, a separate analysis was
performed in which the efficiencies for neutrons with the same energy but different
positions were determined. These results provided correction factors for the data
obtained at charged particle angles of 24$^\circ$ and 30$^\circ$. The additional
systematic error introduced by this procedure will be discussed later.

\subsection{Charged Particle Detection Efficiency}

Recoiling charged particles will lose varying amounts of energy in the
target depending on the location of the scattering event, and then will
suffer further energy losses in the windows, air, and $\Delta E$
scintillator before being detected in the CsI.  For $nd$ scattering, a
Monte Carlo simulation of these effects showed that all deuterons produced
with energies greater than 35 MeV are detected.  Since the lowest recoil
deuteron energy in the present work is about 40 MeV, the efficiency of
charged particle detection is assumed to be unity.

\subsection{Neutron-deuteron Scattering Analysis}\label{sec:nd}

\begin{figure}[hbt!]
  \centering
  \includegraphics[scale=0.8]{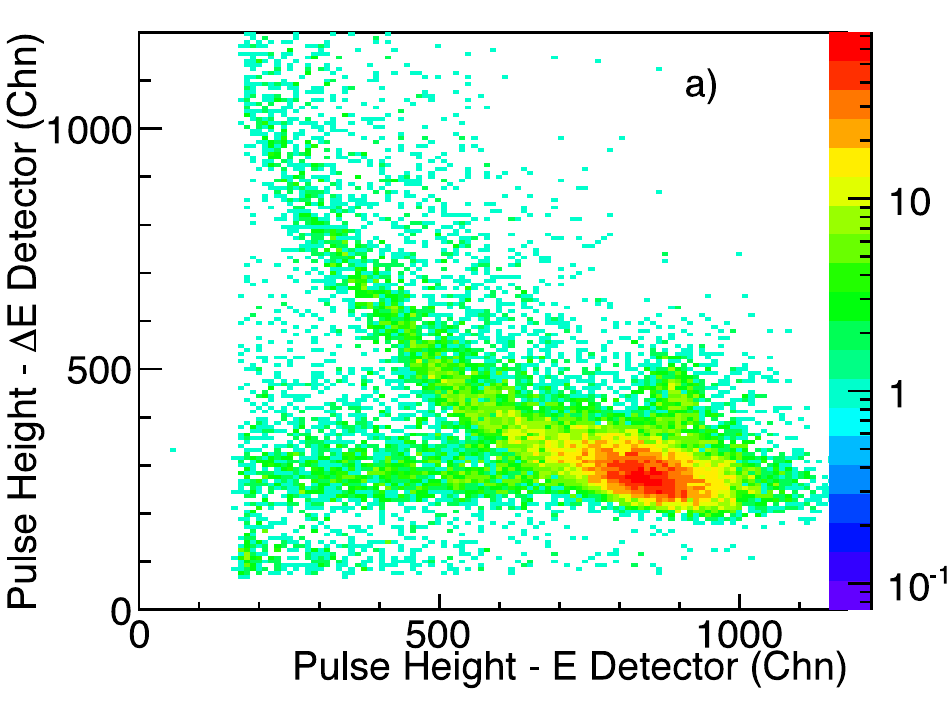}
  \includegraphics[scale=0.8]{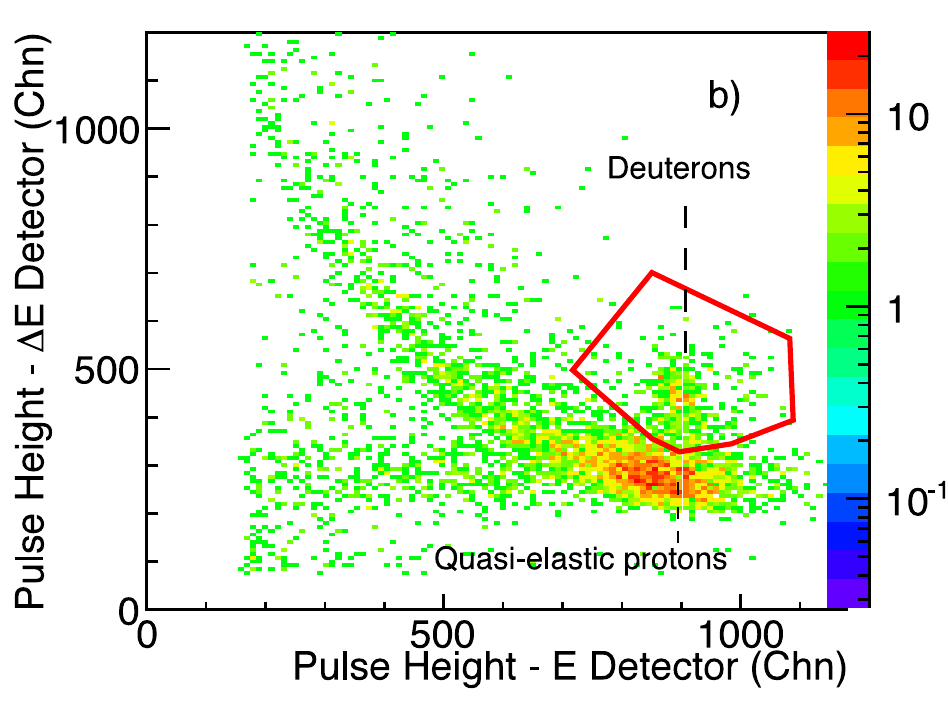}
  \caption{$\Delta E-E$ histograms of $nd$ events for deuteron recoil angle
  $\theta_d$ = 36$^{\circ}$ and incident neutron energy $T_\mathrm{{beam}}=210$~MeV a) before,
  b) after the cuts on the neutron timing as described in the text. }\label{fig:EdE_nd}
\end{figure}

Analysis of the $nd$ data was more challenging than that of the $np$ data,
since the $nd$ elastic scattering cross section is much smaller than the
$d(n,np)n$ quasi-elastic scattering cross section. Although in principle
detection of a deuteron (in singles mode)in a charged-particle telescope
is a unique signature of elastic scattering, in practice the limited resolution
of these detectors does not allow unambiguous separation of deuterons and protons.
This difficulty was resolved by employing the neutron-charged particle coincidence
trigger. In coincidence mode, successive cuts were applied on both the difference
between and the sum of the arrival times of the pulses at the left and right ends
of the neutron bars. The difference provides the location on the bar and the sum
the total time-of-flight of the neutrons. The elastically scattered neutrons
associated with a given recoil deuteron angle satisfy kinematic conditions with
respect to both position and time-of-flight, whereas the values for background
events are distributed over the possible range. Typical $\Delta E-E$ histograms
both before and after these conditions were imposed are shown in
Figure~\ref{fig:EdE_nd}. The suppression of the quasi-elastic events now allows
the recoil deuteron peak to be observed.

\begin{figure}[hbt!]
  \centering
  \includegraphics[width=8.5cm]{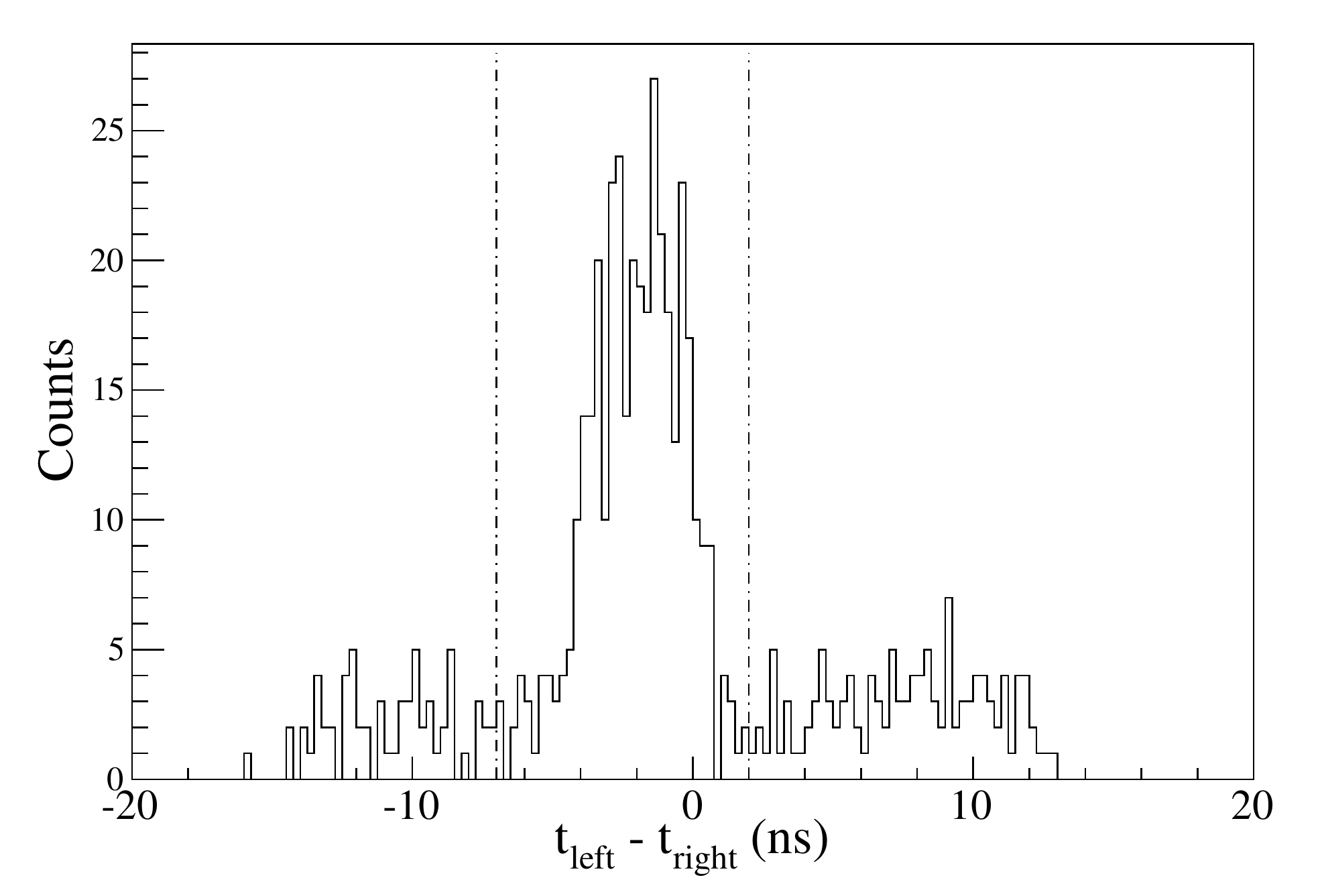}
  \caption{The neutron-bar time difference histogram at $\theta_d = 36^{\circ}$
  and $T_\mathrm{{beam}}$= 170~MeV. }\label{fig:nbar2_170}
\end{figure}

A typical measured distribution of neutron scattering angles for neutrons in
coincidence with deuterons satisfying the particle-identification cut is shown
in Figure~\ref{fig:nbar2_170}. In this histogram the neutron scattering angle
is represented by the measured difference between the pulse arrival times at the
left and right ends of the scintillator bars. Neutrons from the elastic $nd$
events are easily recognizable above the background. The region of interest is
selected, and the events in that region are counted. For the majority of cases
the background is relatively constant, allowing it to be easily interpolated
under the region of the elastic peak and then subtracted from the total peak
yield. In a few cases the background was not uniform, due to the residual
quasi-elastic protons which remained inside the deuteron cut. For these cases,
a mathematical model was used to estimate the background. This procedure
introduced additional systematic errors, estimated to be around 10\%. 

A correction factor was applied to each measurement to account for a slight
degree of polarization in the neutron beam. The neutron polarization on the
15R flight path was measured previously with a CH$_2$ polarimeter using left-right
symmetric detection of $np$ scattering yields \cite{houg_2001}. The average
vertical component of the beam polarization was found to be about 0.06, and
to vary slightly with energy. The correction factor to the measured yield is
given by $1 - p_n * A_y$, where $p_n$ represents the polarization component
normal to the scattering plane, and $A_y$ represents the analyzing power for
the reaction. This correction was applied to both the $np$ and the $nd$
scattering data. The same partial wave analyses used in the evaluation of
$np$ cross sections were also used to determine $np$ elastic analyzing powers
to a precision of about 0.01. The estimated $nd$ elastic analyzing powers
had a greater uncertainty, since they were based on a limited set of available
data. Because of the small size of $p_n$, this uncertainty in $A_y$ constitutes
a relatively small systematic uncertainty, even if no correction is made.
In the end, the corrections due to polarization are about 1\%, and are not
significant compared with the other errors.

\subsection{Systematic Errors}

An error of about 3\% was assigned due to small deviations of the detectors from
their assumed positions which was simulated with a Monte Carlo study. An error of
5\% was due to $np$ normalization. As mentioned earlier, the normalization coefficient
may not be the same for LH${}_2$ and LD${}_2$ target due to different boiling points
and bubbling levels. Therefore, a generous error of 10\% was introduced as there is
no way of measuring this effect. The neutron detection efficiency error is about 5\%
on the average, varying slightly with energy. A separate analysis showed that the
error due to the unknown uranium cross section has to be included in addition to the
$np$ normalization error. The systematic error of interpolation in the unknown
cross section region is estimated to be 9\%.

Additional errors were introduced for the most backward-angle data. The position
dependence in the neutron detection efficiency led to systematic uncertainties of
14\% at 24$^{\circ}$, and 11\% at 30$^{\circ}$. Moreover, the method used to deal
with nonuniform background in the neutron time-of flight difference spectrum at
24$^{\circ}$ introduces about 10\% systematic error. Therefore, the results at
36$^{\circ}$, 42$^{\circ}$ and 48$^{\circ}$ carry about 16\% total systematic
uncertainty, whereas the results at 24$^{\circ}$ and 30$^{\circ}$ have about 24\%
and 19\% total systematic error, respectively.

\section{Results \label{res}}

\begin{figure*}[hbt!]
  \centering
  \includegraphics[width=\textwidth]{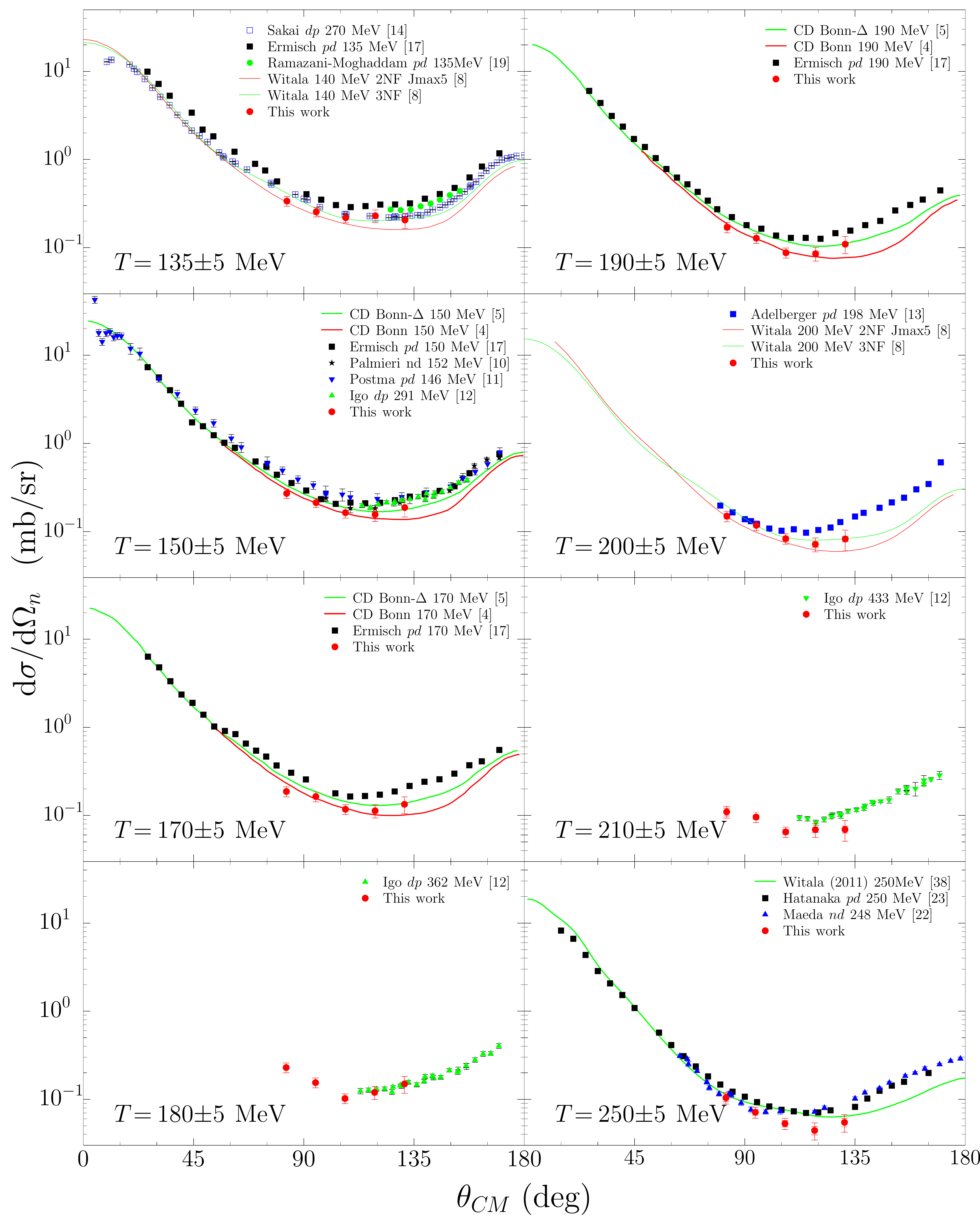}
  \caption{The cross section results. The experimental data and the theoretical work
  prior to this study is taken from \cite{bib:Sakai_2000, Ermi_2005, bib:Ramazani2008,
  bib:Palmieri1972, bib:Postma1961, Igo_1972, PhysRevD.5.2139, Hata_2002,
  Maed_2007} and \cite{Wita_1998, bib:machleidt2001, bib:deltuva2003, bib:witala2011},
  respectively. }
  \label{fig:nd_cs1}
\end{figure*}

Cross sections for elastic $nd$ scattering were obtained over the range of neutron
(lab) energies from 130 to 255 MeV. At lower energies, the energy loss of the
recoil deuterons in the target and air was found to be too large; at energies 
above 255 MeV, particle identification based on the $\Delta E$--$E$ plot became
unreliable. The data were binned into $\pm$ 5~MeV intervals at each of a total of
eight centroid beam energies, and the differential cross sections were then calculated
for each of the recoil deuteron angles. These results, and their uncertainties, are
listed in Table~\ref{tab:csRes1}, and are shown as angular
distributions in Figure~\ref{fig:nd_cs1}. The broad energy
coverage apparent in these figures is a unique feature of the present experiment,
while the angular range of these new data is observed to fall in all cases within
the region of the minimum of the cross section -- i.e., where the sensitivity to
the effects of three-nucleon forces is expected to be largest.

Also included in these figures are the relevant data from previous $nd$, $pd$, and
$dp$ scattering experiments. The most extensive comparison that can be made with previous work
is with the KVI $pd$ measurements of references~\cite{Ermi_2003, Ermi_2005}. In nearly
all cases these cross sections are larger than the present results. At our lowest energy,
135 MeV, the current results are in good agreement with the RIKEN $dp$ cross
sections~\cite{bib:Sakai_2000}.

Figure~\ref{fig:nd_cs1} also displays the calculated cross sections. At 135 and 200 MeV, the
Faddeev calculations of reference~\cite{Wita_1998,Coon_1981} are shown both with and
without the inclusion of 3NF. As expected, these results are significantly different
from each other only at angles larger than about 90$^{\circ}$. At both of these 
energies, the present cross sections are found to be in better agreement with the 3NF
results. At 150, 170, and 190 MeV the data are compared with calculations based on the
CD Bonn potential without and with the inclusion of a $\Delta$--$N$
component~\cite{bib:deltuva2003}. At the largest angle where this effect is most
significant, the data favor its inclusion.

{
\linespread{1}

\begin{table*}[h!]
\caption{The $nd$ cross section results }\label{tab:csRes1}
\centering
\vspace{3mm}
\begin{tabular}{| c | c | c | c  c  c  c  c |}
\hline
  $T_\mathrm{beam}$ & $\theta_d$ Lab & $\theta_n$ CM & $d\sigma / d\Omega_n$  & $\pm$  & $\epsilon_\mathrm{stat}$  & $\pm$ &
$\epsilon_\mathrm{sys}$  \\
(MeV) &\hspace*{0mm}(deg)\hspace*{0mm} & \hspace*{0mm}(deg)\hspace*{0mm} & \hspace*{1cm}($\mu$b/sr)\hspace*{-1cm} &&& & \\
\hline\hline
\multirow{5}{*}{$135\pm 5$}
&  24  &  131  &  206  &  $\pm$  &  6  &  $\pm$  &  45  \\
&  30  &  119  &  230  &  $\pm$  &  6  &  $\pm$  &  44  \\
&  36  &  107  &  218  &  $\pm$  &  5  &  $\pm$  &  34  \\
&  42  &   95  &  255  &  $\pm$  &  7  &  $\pm$  &  40  \\
&  48  &   83  &  337  &  $\pm$  & 14  &  $\pm$  &  53  \\
\hline
\multirow{5}{*}{$150\pm 5$}
&  24  &  131  &  187  &  $\pm$  &  6  &  $\pm$  &  45  \\
&  30  &  119  &  156  &  $\pm$  &  5  &  $\pm$  &  29  \\
&  36  &  107  &  164  &  $\pm$  &  5  &  $\pm$  &  27  \\
&  42  &   95  &  213  &  $\pm$  &  6  &  $\pm$  &  34  \\
&  48  &   83  &  270  &  $\pm$  &  9  &  $\pm$  &  42  \\
\hline
\multirow{5}{*}{$160\pm 5$}
&  24  &  131  &  143  &  $\pm$  &  5  &  $\pm$  &  33  \\ 
&  30  &  119  &  161  &  $\pm$  &  5  &  $\pm$  &  31  \\ 
&  36  &  107  &  142  &  $\pm$  &  5  &  $\pm$  &  22  \\ 
&  42  &   95  &  209  &  $\pm$  &  7  &  $\pm$  &  33  \\ 
&  48  &   83  &  301  &  $\pm$  &  9  &  $\pm$  &  48  \\ 
\hline
\multirow{5}{*}{$170\pm 5$}
&  24  &  131  &  134  &  $\pm$  &  6  &  $\pm$  &  32  \\ 
&  30  &  119  &  113  &  $\pm$  &  5  &  $\pm$  &  21  \\ 
&  36  &  107  &  117  &  $\pm$  &  5  &  $\pm$  &  18  \\ 
&  42  &   95  &  164  &  $\pm$  &  6  &  $\pm$  &  26  \\ 
&  48  &   83  &  187  &  $\pm$  &  8  &  $\pm$  &  30  \\ 
\hline
\multirow{5}{*}{$180\pm 5$}
&  24  &  131  &  149  &  $\pm$  &  7  &  $\pm$  &  34  \\ 
&  30  &  119  &  120  &  $\pm$  &  6  &  $\pm$  &  22  \\ 
&  36  &  107  &  102  &  $\pm$  &  5  &  $\pm$  &  16  \\ 
&  42  &   95  &  155  &  $\pm$  &  6  &  $\pm$  &  25  \\ 
&  48  &   83  &  229  &  $\pm$  &  9  &  $\pm$  &  36  \\ 
\hline   
\multirow{5}{*}{$190\pm 5$}
&  24  &  131  &  109  &  $\pm$  &  6  &  $\pm$  &  26  \\ 
&  30  &  119  &   85  &  $\pm$  &  5  &  $\pm$  &  16  \\ 
&  36  &  107  &   87  &  $\pm$  &  4  &  $\pm$  &  14  \\ 
&  42  &   95  &  128  &  $\pm$  &  6  &  $\pm$  &  20  \\ 
&  48  &   83  &  170  &  $\pm$  & 12  &  $\pm$  &  26  \\ 
\hline   
\end{tabular}\hspace{5mm}
\begin{tabular}{| c | c | c | c  c  c  c  c |}
\hline
  $T_\mathrm{beam}$ & $\theta_d$ Lab & $\theta_n$ CM & $d\sigma / d\Omega_n$  & $\pm$  & $\epsilon_\mathrm{stat}$  & $\pm$ &
$\epsilon_\mathrm{sys}$  \\
(MeV) &\hspace*{0mm}(deg)\hspace*{0mm} & \hspace*{0mm}(deg)\hspace*{0mm} & \hspace*{1cm}($\mu$b/sr)\hspace*{-1cm} &&& & \\
\hline\hline
\multirow{5}{*}{$200\pm 5$}
&  24  &  131  &   82  &  $\pm$  &  6  &  $\pm$  &  23  \\ 
&  30  &  119  &   72  &  $\pm$  &  5  &  $\pm$  &  14  \\ 
&  36  &  107  &   83  &  $\pm$  &  5  &  $\pm$  &  13  \\ 
&  42  &   95  &  117  &  $\pm$  &  6  &  $\pm$  &  18  \\ 
&  48  &   83  &  149  &  $\pm$  & 11  &  $\pm$  &  23  \\ 
\hline   
\multirow{5}{*}{$210\pm 5$}
&  24  &  131  &   69  &  $\pm$  &  6  &  $\pm$  &  18  \\ 
&  30  &  119  &   69  &  $\pm$  &  5  &  $\pm$  &  13  \\ 
&  36  &  107  &   65  &  $\pm$  &  4  &  $\pm$  &  10  \\ 
&  42  &   95  &   96  &  $\pm$  &  6  &  $\pm$  &  15  \\ 
&  48  &   83  &  110  &  $\pm$  & 11  &  $\pm$  &  17  \\ 
\hline
\multirow{5}{*}{$220\pm 5$}
&  24  &  131  &   60  &  $\pm$  &  6  &  $\pm$  &  16  \\ 
&  30  &  119  &   78  &  $\pm$  &  5  &  $\pm$  &  15  \\ 
&  36  &  107  &   86  &  $\pm$  &  4  &  $\pm$  &  13  \\ 
&  42  &   95  &  116  &  $\pm$  &  6  &  $\pm$  &  18  \\ 
&  48  &   83  &  211  &  $\pm$  &  9  &  $\pm$  &  33  \\ 
\hline
\multirow{5}{*}{$230\pm 5$}
&  24  &  131  &   64  &  $\pm$  &  6  &  $\pm$  &  16  \\ 
&  30  &  119  &   51  &  $\pm$  &  6  &  $\pm$  &  10  \\ 
&  36  &  107  &   69  &  $\pm$  &  6  &  $\pm$  &  11  \\ 
&  42  &   95  &   77  &  $\pm$  &  7  &  $\pm$  &  12  \\ 
&  48  &   83  &   93  &  $\pm$  & 11  &  $\pm$  &  14  \\ 
\hline   
\multirow{5}{*}{$240\pm 5$}
&  24  &  131  &   40  &  $\pm$  &  6  &  $\pm$  &  12  \\ 
&  30  &  119  &   42  &  $\pm$  &  5  &  $\pm$  &   8  \\ 
&  36  &  107  &   52  &  $\pm$  &  4  &  $\pm$  &   8  \\ 
&  42  &   94  &   90  &  $\pm$  &  5  &  $\pm$  &  14  \\ 
&  48  &   82  &  127  &  $\pm$  &  8  &  $\pm$  &  20  \\ 
\hline
\multirow{5}{*}{$250\pm 5$}
&  24  &  131  &   55  &  $\pm$  &  6  &  $\pm$  &  12  \\ 
&  30  &  119  &   44  &  $\pm$  &  5  &  $\pm$  &   8  \\ 
&  36  &  106  &   53  &  $\pm$  &  4  &  $\pm$  &   8  \\ 
&  42  &   94  &   71  &  $\pm$  &  5  &  $\pm$  &  11  \\ 
&  48  &   82  &  104  &  $\pm$  & 10  &  $\pm$  &  16  \\ 
\hline
\end{tabular}
\end{table*}
}

\section{Conclusion \label{conc}}

In this study, the differential cross section for $nd$ elastic scattering was
measured in a continuous incident neutron energy range from 135 to 250 MeV, by
detecting scattered neutrons and recoil deuterons in coincidence, with the aim
of elucidating the contribution of three-nucleon forces (3NF), in particular the
energy dependence of this effect. The absolute scale of the $nd$ cross section
was determined by concurrent measurements of $np$ elastic scattering, which were
also used to determine the neutron detection efficiency.

The effect of 3NF is clearly seen in this work. The data at angles near the
minimum in the cross section, where the 3NF contribution is most effective,
are in excellent agreement at all energies with the theoretical predictions.

The 3NF effect could be further tested by confronting the present data with
theoretical predictions for the differential cross section at fixed 
angles as a function of incident neutron energy.

\begin{acknowledgments}
We wish to acknowledge K. Boddy for her contributions to various aspects of
this experiment. We thank the staff of the Los Alamos Neutron Science Center
for reliable delivery of beam and for help in preparing the experimental area,
particularly in maintaining the liquid deuterium target. This work was supported
in part by the U.S. Department of Energy and the National Science Foundation,
the Scientific and Technological Research Council of Turkey (107T538), and
Bogazici University Research Fund (BAP6057).
\end{acknowledgments}

\bibliography{nd}

\end{document}